\documentclass{article}
\usepackage{graphicx}
\begin{document}
\title{Can measuring entanglement be easy?}
\author{S.J. van Enk\\
Bell Labs, Lucent Technologies\\
600-700 Mountain Ave,
Murray Hill, NJ 07974}
\maketitle
\begin{abstract}
I raise some doubts concerning a protocol recently applied in an experiment 
\cite{walborn}
to measure entanglement.  The protocol is much simpler than other known entanglement-verification methods, but, I argue, needs assumptions (namely that the state generated is known and pure) that are too strong to be allowed and that are not justified in most experiments. An extension of the protocol suggested in \cite{flo}
is much harder to implement and still relies on assumptions not justified in entanglement-verification protocols, as demonstrated by an explicit example.
\end{abstract}
In Ref.~\cite{walborn} an experiment is described where the concurrence of an entangled state is measured in
a particularly simple way. One motivation for the procedure is that methods such as entanglement witnesses work only for certain entangled states and fail (i.e., give a negative answer) for others.
However, I argue here that the method used in \cite{walborn}, although based on a correct theoretical proposal \cite{mintert}, works only for a special type of states and, what is worse, gives a false positive answer for other states. 
Moreover, the other point presented in \cite{walborn} in favor of their protocol, namely that more complicated measurements that would be needed for quantum tomography or state reconstruction
are not necessary, is moot, as the only reason for this is that the state to be tested is assumed to be {\em known}.

An extension of the method discussed in Ref.~\cite{flo} (see below) will not give false positive answers, but will overestimate entanglement for a generic class of states. Moreover, that extension will require much more complicated measurements, nullifying one of the arguments in favor of the simple protocol of \cite{walborn}.

The main objection to the protocol of \cite{walborn} {\em as entanglement-verification protocol}, is that it only
works if the state to be tested is known and pure. The protocol cannot, therefore, be regarded as an independent check on entanglement. 
For example, that disqualifies the protocol as a test to be used in quantum key distribution protocols such as the entanglement-based version of BB84, where the provable presence of entanglement is a precondition for security of the protocol \cite{norbert}. In the cryptographic setting it is obvious the legimate users of the protocol cannot rely on any assumptions about the form of the state they share. The new method proposed in \cite{flo} suffers from a similar, albeit milder, drawback, namely it still assumes too much of the states to be tested. An explicit example is given when discussing the third objection to the experiment \cite{walborn}, see below.

The method of \cite{walborn} is based on the following observation. Suppose Alice and Bob each have two quantum systems (for simplicity taken to be qubits from here on) denoted by $A_k$ for $k=1,2$ for Alice and $B_k$ for Bob. Suppose further they ascribe two {\em identical} pure states $|\Psi\rangle_1$
and $|\Psi\rangle_2$
to the bipartite systems $A_1, B_1$ and $A_2, B_2$. 
If those states are product states of the form 
\begin{equation}
	|\Psi\rangle_k=|x\rangle_{A_k}\otimes |y\rangle_{B_k},
\end{equation}
then the state of the two systems $A_1$ and $A_2$ on Alice's side lives in the symmetric subspace,
and the same holds for Bob's state on his systems $B_1$ and $B_2$.
Hence, denoting the probabilities to project onto the respective antisymmetric subspaces of Alice's and Bob's qubits by $P_a^{A,B}$, we would have $P_a^A=P_a^B=0$.
And so Ref.~\cite{mintert} correctly concludes a nonzero value for $P_a^A$ (or for $P_a^B$ of course) implies entanglement.  Moreover it is shown that the concurrence $C$ is in fact given by 
\begin{equation}\label{C}
C=2\sqrt{P_a^A}.	
\end{equation}
So far the theory.

The experimental procedure of \cite{walborn} now consists of generating two bipartite states, measuring $P_a^A$, assuming the two bipartite states are pure and identical, and converting $P_a^A$ to a measure of entanglement through the relation (\ref{C}).
The result reported is that the experimentally determined value of the concurrence reached the maximum (for a 2-qubit state) of 1. 

Here are three objections to the experiment:
{\bf First}, finding $P_a^A$ to be nonzero in an actual experiment may just as well be interpreted
	as indicating that the two copies 1 and 2 are {\em not} in the same pure state. 
Since	the experiment {\em only} measured $P_a^A$, I would say nothing conclusive can be concluded about entanglement. One would need more measurements (such as quantum tomographic measurements) to actually confirm the two bipartite systems are in the same pure state. But those measurements will be much harder to implement and thus defeat the main purpose of \cite{walborn}. 

{\bf Second}, even if one believes, without doing any additional measurements, that systems 1 and 2 are in the same state, measuring $P_a^A$ is not sufficient to demonstrate entanglement.
Namely,	since the state generated in the actual experiment is not known (otherwise there would be no need for any measurement),	we may use, inspired by the Quantum De Finetti Theorem \cite{finetti}, the more careful {\em Ansatz} that ``being in the same but unknown state'' may mean\footnote{Without more conditions one cannot conclude the state of two ``identical copies'' {\em must} be of the De Finetti form \cite{finetti}. Here it is used only as one of several alternative state descriptions one may use.}
\begin{equation}\label{dF}
	\rho_{1,2}=\sum_i p_i \rho^{(i)}_1 \otimes
	\rho^{(i)}_2,
\end{equation}
where we have a mixture of mixed states $\rho^{(i)}$, assumed identical for systems 1 and 2.
Indeed, this form (\ref{dF}) is consistent with {\em all} measurement statistics on 
systems 1 and 2 being the same. But assuming the form (\ref{dF}) rather than a product of pure states $|\Psi\rangle_1\otimes|\Psi\rangle_2$
invalidates the argument for $P_a$ being a measure of entanglement.
For instance, it is easy to check that assuming both copies to be in the completely mixed state still leaves a probability $P_a^A=P_a^B=1/4$ to be projected onto the antisymmetric subspace, which would lead one to believe there is maximal entanglement ($C=1$), although there is none.

{\bf Third}, let us assume that one has strong reasons to believe that 
the joint state of the two copies $1$ and $2$ is of the form 
\begin{equation}\label{pure}
	\rho_{1,2}=\sum_i p_i |\Psi^{(i)}\rangle_1\langle \Psi^{(i)}| \otimes
	|\Psi^{(i)}\rangle_2\langle \Psi^{(i)}|,
\end{equation}
where the states $|\Psi^{(i)}\rangle_k$ are {\em pure} states on a Hilbert space $H^{A_k}\otimes H^{B_k}$.
That is, in the De Finetti form (\ref{dF}) we assume that all possible density matrices actually correspond to {\em pure} states. 
This situation corresponds to the statement ``the two bipartite systems 1 and 2 are in the same pure state, we just do not know which one.''
In this situation the single-copy concurrence could be calculated by taking the infimum
over all possible decompositions of the single-copy density matrix of the average concurrence.
But the experiment does not measure or calculate an infimum, it just measures $P_a^A$. This quantity can be written as
\begin{equation}\label{pureC}
	P_a^A=\big[\sum_i p_i C(|\Psi^{(i)}\rangle)\big]^2/4,
\end{equation}
by inverting relation (\ref{C}).
However, using this value of $P_a^A$ to calculate the concurrence through (\ref{C}) 
will in general overestimate the entanglement. The reason is that using the expression (\ref{pureC}) boils down to using the particular decomposition
of the single-copy density matrix
\begin{equation}\label{deco}
	\rho_1=\sum_i p_i |\Psi^{(i)}\rangle \langle \Psi^{(i)}|.
\end{equation}
But this decomposition in general does not coincide with the one that yields the infimum of the average concurrence.
In other words, although the decomposition (\ref{deco}) may seem privileged because of the form (\ref{pure}) of the 2-copy density matrix, it is not the correct decomposition for calculating entanglement.

Here is an explicit example that illustrates this last point and that also illustrates why the new method proposed in \cite{flo}, although an improvement, still fails.
A state of the form (\ref{pure}) featured in discussions about reference frames
in quantum communication \cite{reference}: 
\begin{equation}\label{phase2}
\rho_{1,2}=\int\frac{{\rm d} \phi}{2\pi}
\frac{1}{4}
\big[(|0\rangle|1\rangle+\exp(i\phi)|1\rangle|0\rangle)
(\langle 0|\langle 1|+\exp(-i\phi)\langle 1|\langle 0|)\big]^{\otimes 2}
\end{equation}
In words, we have two copies of a maximally entangled state, {\em but} with the an unknown phase $\phi$ appearing in both copies.
This state is not maximally entangled, and has in fact an amount of entanglement $E=1/2$ \cite{reference}. This is because the mixture over $\phi$ is equivalent to an equal mixture of
the unentangled correlated states $|0\rangle|0\rangle\otimes |1\rangle|1\rangle$
and $|1\rangle|1\rangle\otimes |0\rangle|0\rangle$ on the one hand, and the unambiguously
maximally entangled state $(|0_L\rangle\otimes |1_L\rangle + |1_L\rangle\otimes |0_L\rangle)/\sqrt{2}$,
where the ``logical'' $0$ and $1$ are defined as
\begin{equation}
|0_L\rangle=|0\rangle|1\rangle;\,\,\,
|1_L\rangle=|1\rangle|0\rangle.
\end{equation}
However, measuring $P_a^A$ on such a mixture would give $P_a^A=1/4=P_a^B$ and hence one would incorrectly conclude $C=1$. This overestimates the entanglement by a factor of 4, since 
$C=1$ refers to single-copy entanglement, whereas the correct $E=1/2$ result quoted above refers to the total entanglement in the {\em two} copies. 

Note that if one has {\em many} copies of the same state with unknown phase, then asymptotically one does have 1 ebit of entanglement per copy. This entanglement is in encoded form, and can be extracted by joint measurements on the many copies.
The method of \cite{walborn} does give the right answer for the asymptotic case, but only if it is properly interpreted as refering to the entanglement in the joint state, not the individual states, {\em and} if
one assumes the form (\ref{pure}) instead of (\ref{dF}) generalized to many copies. 
So the assumption in this case would no longer be that two copies need to be in the same known pure state, but {\em many} copies need to be in the same unknown pure state.

As pointed out in \cite{flo}
if one just lets Bob do the same measurement as Alice, then from the correlations
between $P_a^A$ and $P_a^B$ they can indeed check whether the state description (\ref{dF}) is needed (namely when the two measurement outcomes differ sometimes) and when not (namely when those measurement results are always the same).
However, in the example (\ref{phase2}) the new method still overestimates the entanglement. Namely,
even the new method would claim $C=1$ in a state of the form (\ref{phase2}), although the entanglement is only $E=1/2$ (per two copies). Here the difference arises from the assumption that Ref.~\cite{flo} makes,
that the joint state
of the two copies is of the form $\rho\otimes\rho$. But again, since that form is a {\em special} case of the form (\ref{dF}) it has to be verified. The measurements suggested in \cite{flo} do not succeed in doing that.

Moreover, in the quantum cryptographic context an eavesdropper, Eve, can exploit Alice's and Bob's measurements in yet another way. If Alice and Bob perform their measurements on given pairs of entangled states\footnote{In the experiment \cite{walborn} the two copies on each side reside in two degrees of freedom of the same single photon: in that case there is no other practical choice for Alice and Bob but to measure
$P_a$ on those photons}, Eve can simply prepare either antisymmetric states of the two copies on both sides, or symmetric states, to force Alice and Bob always getting the same result $P_a^A=P_a^B$. 

What is the difference between theory and experiment here? In the calculation of entanglement of a pure state of one bipartite system one is free to assume as a theoretical tool the existence of a second system in the same pure state. But experimentally there is no guarantee an actual second system is in the same pure state. One thus would have to do more measurements, especially those that the protocol of \cite{walborn} explicitly tries to avoid. Moreover, there is no guarantee the system is in a pure state. But where theoretically one can calculate the infimum over all possible pure-state decompositions of a mixed density matrix of the average concurrence, the experiment \cite{walborn} just measured $P_a^A$, which amounts to assuming a particular pure-state decomposition of the mixed density matrix. This will always overestimate the entanglement in the state generated.  
Even the improved set of measurements of \cite{flo} fails as an entanglement-verification test in an
adversarial (e.g. quantum cryptographic) setting.

Finally, I thank the authors of \cite{walborn} for useful discussions, especially with Andreas Buchleitner and Florian Mintert.

\end{document}